\documentclass[11pt]{article}
\usepackage{amsfonts,amsmath,amssymb}
\usepackage{mathtools}
\usepackage{xspace}
\usepackage{bm}
\usepackage{braket}
\usepackage{color}
\usepackage[left=2.8cm,right=2.8cm,top=3cm,bottom=4cm]{geometry}
\usepackage{graphicx}
\usepackage{cite}

\DeclareRobustCommand{\ensuremathrm}[1]{\ensuremath{\mathrm{#1}}\xspace}

\DeclareRobustCommand{\rs}{\ensuremathrm{s}}

\DeclareRobustCommand{\jet}{\ensuremathrm{jet}\xspace}
\DeclareRobustCommand{\cut}{\ensuremathrm{cut}\xspace}
\DeclareRobustCommand{\alphas}{\ensuremath{\alpha_\rs}\xspace}
\DeclareRobustCommand{\mur}{\ensuremath{\mu_\mathrm{R}}\xspace}
\DeclareRobustCommand{\muR}{\mur}

\DeclareRobustCommand{\PH}{{\ensuremathrm{H}}\xspace}
\DeclareRobustCommand{\PZ}{{\ensuremathrm{Z}}\xspace}
\DeclareRobustCommand{\Pg}{{\ensuremathrm{g}}\xspace}

\DeclareRobustCommand{\Pep}{{\ensuremathrm{e^+}}\xspace}
\DeclareRobustCommand{\Pem}{{\ensuremathrm{e^-}}\xspace}

\DeclareRobustCommand{\Pq}{{\ensuremath{q}}\xspace}
\DeclareRobustCommand{\Paq}{{\ensuremath{\bar{q}}}\xspace}

\DeclareRobustCommand{\NNLOJET}{\textsc{NNLOjet}}

\begin{document}
\unitlength1cm
\begin{titlepage}
\vspace*{-1cm}
\begin{flushright}
IPPP/17/57, ZU-TH 22/17
\end{flushright}
\vskip 3.5cm

\begin{center}
{\Large\bfseries\boldmath NNLO QCD corrections to \\[2mm] event orientation in $\Pep\Pem$ annihilation}
\vskip 1.cm
{\large T.~Gehrmann$^a$, E.W.N.~Glover$^b$, A.~Huss$^c$, J.~Niehues$^a$, H.~Zhang$^a$}
\vskip .7cm
{\it
$^a$ Department of Physics, Universit\"at Z\"urich,
Winterthurerstrasse 190,\\ CH-8057 Z\"urich, Switzerland
}\\[2mm]
{\it
$^b$ Institute for Particle Physics 
Phenomenology, University of Durham, Durham DH1 3LE, UK}\\[2mm]
{\it $^c$ Institute for Theoretical Physics, ETH, CH-8093 Z\"urich, Switzerland}\\[2mm]
\end{center}
\vskip 2cm

\begin{abstract}
We present a new implementation of the NNLO QCD corrections to three-jet final states and related 
event-shape observables in electron--positron annihilation. Our implementation is based on the 
antenna subtraction method, and is performed in the \NNLOJET framework. The calculation improves 
upon earlier results by taking into account the 
full kinematical information on the initial state momenta, thereby allowing the event orientation 
to be computed to NNLO accuracy. 
We find the event-orientation distributions at LEP and SLC to be 
very robust under higher order QCD corrections. 
\end{abstract}
\end{titlepage}
\newpage

The production of hadronic final states in electron--positron annihilation at high energies offers 
a unique laboratory for testing the theory of the strong interaction, quantum chromodynamics (QCD). 
Experiments at LEP and SLD have collected a wealth of precision data on jet cross sections and event-shape 
distributions~\cite{alephqcd,opalqcd,l3qcd,delphiqcd,sldqcd}. Precision studies of these data included 
establishing the gauge group structure of QCD, measurements of the strong coupling constant and  
investigations of the all-order structure of large logarithmic effects in QCD~\cite{qcdbook}.
These studies rely on the comparison between the data and theory predictions, with the 
inherent uncertainty of the theoretical calculations due to truncating a perturbative expansion 
often being a limiting factor. Most of the original LEP and SLD studies were based on the then
available
NLO theory predictions for event shapes and cross sections~\cite{ert,kn,fourjet}. 
These calculations are in the form of fixed order parton-level codes, which produce weighted events containing sets of parton momenta and which can adapt in a flexible manner to the jet definition and event-shape variables 
used in the experimental studies. 

The calculation of NNLO corrections to three-jet production and related event-shape 
observables~\cite{our3j,weinzierl3j} enabled these data to be confronted with 
increasingly precise predictions, and led to 
a variety of new precision QCD studies~\cite{eeas}.
The calculation of jet-like observables at NNLO requires a method for
the cancellation of infrared singular contributions across channels of different 
partonic multiplicity.  
Both early calculations~\cite{our3j,weinzierl3j} 
(with the EERAD3 code of~\cite{our3j} documented in detail in its public release~\cite{eerad3}) 
were based on the antenna subtraction 
method~\cite{ourant}. They have been recently complemented by a new calculation~\cite{trocsanyi3j}
based on the colourful-subtraction method~\cite{colorful}. 

To apply the antenna subtraction method to a broad number of processes, we are currently 
developing the \NNLOJET code, which is a fixed order parton-level code that provides the framework for the implementation 
of jet production to NNLO accuracy.  Besides 
containing the necessary event generation infrastructure (phase-space integration, event handling and analysis routines), it 
supplies the unintegrated and integrated antenna functions and the phase-space mappings relevant to all kinematical 
situations. The multi-dimensional phase space integration is performed using the adaptive Monte Carlo 
integrator VEGAS~\cite{vegas}. Processes included in \NNLOJET  up to now 
are $\PZ$ and $\PZ+\jet$ production~\cite{ourzj}, $\PH$ and $\PH+\jet$ production~\cite{ourhj} as well as 
 single-inclusive and di-jet production in hadron--hadron collisions~\cite{2jnew} and in 
 deep inelastic scattering~\cite{ourdisj}.

Our new implementation 
of the NNLO QCD corrections to  $\Pep\Pem\to 3j$ is performed in the \NNLOJET framework. The 
relevant matrix elements correspond to different kinematical crossings of the ones already used~\cite{meRR,meRV,meVV} 
in the $\PZ+j$ and deeply inelastic jet production processes. The structure of the antenna subtraction terms 
for these matrix elements is documented in detail in~\cite{our3jtech}. We validated the new implementation against 
EERAD3~\cite{eerad3} for the canonical set 
of LEP event shapes and jet cross sections. 
While the EERAD3 implementation~\cite{eerad3} was based on the
matrix elements for virtual photon decay $\gamma^*\to \Pq\Paq\Pg$ (and higher order corrections to it), \NNLOJET now contains the 
full $\Pep\Pem\to \Pq\Paq\Pg$ matrix elements through to NNLO. 
It therefore allows to properly account for the correlation between the final-state parton directions and the incoming electron and positron beams.

Most of the  LEP and SLC measurements of event shapes and jet cross sections~\cite{alephqcd,opalqcd,l3qcd,delphiqcd,sldqcd} were corrected to 
a full $4\pi$ acceptance. They do not depend on the angular correlation between the final state hadrons and the incoming electron--positron 
direction. Measurements of fiducial cross sections (restricted to the actual acceptance of the detector) are typically not available, 
the only exceptions being a few studies of oriented event-shape distributions~\cite{opalor,delphior}, which are measured 
in fixed intervals in the angle between the event's thrust axis  and the incoming beam direction. 
An indication of the quality of the extrapolation to full $4\pi$ acceptance can however be gained from studying event-orientation variables, which describe the 
full angular correlation between the hadronic final state and the incoming beams. 

\begin{figure}[t]
  \centering
  \includegraphics[width=0.5\textwidth]{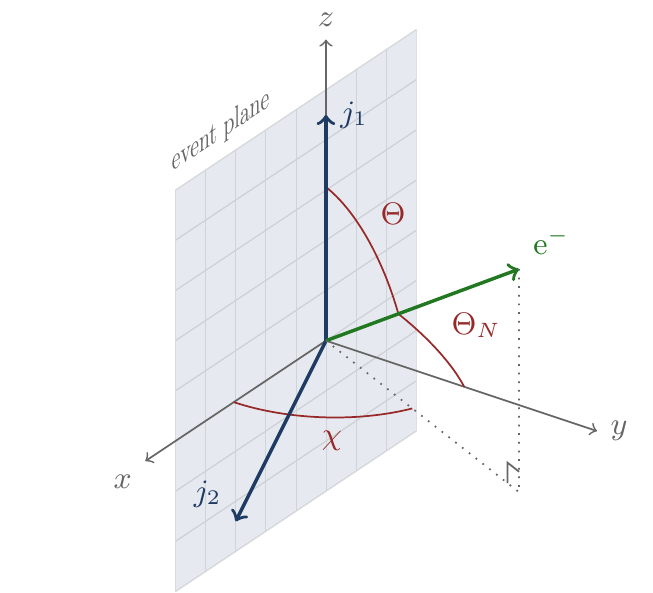}
  \caption{Definition of the three Euler angles characterising the event orientation. 
  $j_1$ denotes the highest-energy jet, $j_2$ the sub-leading jet~\cite{sld}.}
  \label{fig:vardef}
\end{figure}
Three-particle (or three-jet) production in the $\Pep\Pem$ centre-of-momentum frame always results in a final state with momenta in a plane, 
due to momentum conservation. 
The orientation of this event plane with respect to the initial state is described by three Euler angles: 
$(\Theta, \Theta_N,\chi)$~\cite{hoyer12}.  
Taking the event plane in $(x,z)$ and using the 
highest-energy final state object to define the $z$-axis, the incoming electron direction is defined through 
the polar angle $\Theta$ and the azimuthal angle $\chi$. The third angle $\Theta_N$ is then formed by the electron direction 
and the event plane normal. The choice of coordinate system and the definition of the angles is 
displayed in Figure~\ref{fig:vardef}, reproduced from~\cite{sld}.

For three-jet final states, event orientation distributions were measured initially by TASSO~\cite{tasso} and subsequently by 
DELPHI~\cite{delphi}, L3~\cite{l3}, and SLD~\cite{sld}. In all measurements, the JADE algorithm was used to identify 
the final-state jets, and one-dimensional distributions in $\Theta$, $\Theta_N$ or $\chi$
were measured. These measurements were compared with the leading-order, leading-logarithmic 
multi-purpose event generator simulations HERWIG~\cite{herwig} and JETSET/PYTHIA~\cite{jetset}, which all 
provided a very good description of the data. This observation motivates the use of these 
simulation programs to extrapolate the canonical event shape and jet cross sections measurements 
to full $4\pi$ acceptance. 

For this procedure to be reliable, it is however vital that the shapes of the leading order event-orientation 
distributions are not distorted by higher-order QCD corrections. Surprisingly enough, this issue has never been 
investigated in a systematic manner. By using an approximation to the real-radiation contributions,  
NLO QCD corrections to event orientation were estimated to be small in~\cite{schuler}. 
Comparing the JETSET predictions with exact real-radiation matrix elements and parton-shower approximation, SLD~\cite{sld}
attempted to quantify the potential magnitude of real-radiation effects at NLO, which were found to be of limited impact. 

With the \NNLOJET implementation of jet production in $\Pep\Pem$ annihilation, we are now able to compute the 
NLO and NNLO corrections to the event orientation distributions. We consider the kinematical situations that were investigated 
by L3~\cite{l3} and SLD~\cite{sld}, which provide more precise measurements than in the earlier studies. Both experiments 
perform their measurements on an exclusive three-jet sample. The jets are identified using the JADE algorithm~\cite{jade}, 
with a range of jet resolution parameters $y_{\cut}$ for L3, and for fixed $y_{\cut}=0.02$ for SLD. The distributions in 
$(\Theta, \Theta_N,\chi)$ are normalised to the three-jet cross section, such that they all integrate to unity by construction. 
Besides cancelling potential sources of systematic uncertainty, this normalisation condition makes the 
theoretical predictions at leading order independent of $\alphas$. Consequently, the variation of the renormalisation scale will 
not necessarily be a good quantifier 
for the potential impact of higher order corrections, and one should rather look order-by-order into the relative size of the corrections.

The experimental data have all been corrected to $4\pi$ acceptance, 
with SLD~\cite{sld} also providing the uncorrected data. By comparison,
it can be seen that the corrections affect the event orientation distributions only for $\cos(\Theta)\gtrsim 0.7$, $\cos(\Theta_N)\lesssim 0.7$, 
$\chi\lesssim \pi/4$. These  can be identified from Figure~\ref{fig:vardef} as the regions where the event plane comes close to the beam 
direction, such that the final state particles can be partly outside the detector coverage. 

 \begin{figure}[t]
\centering
\includegraphics[width=7cm]{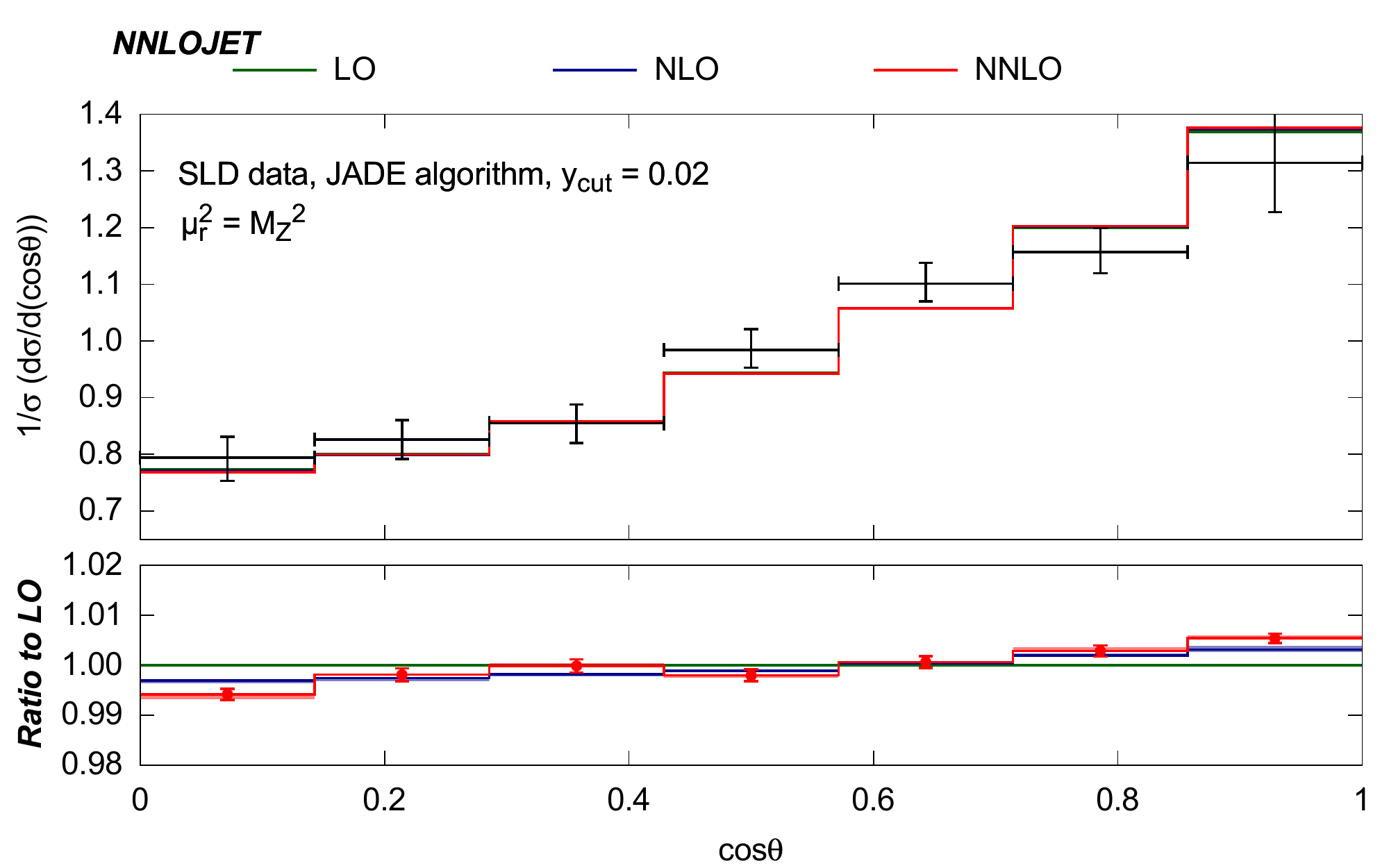}\quad
  \includegraphics[width=7cm]{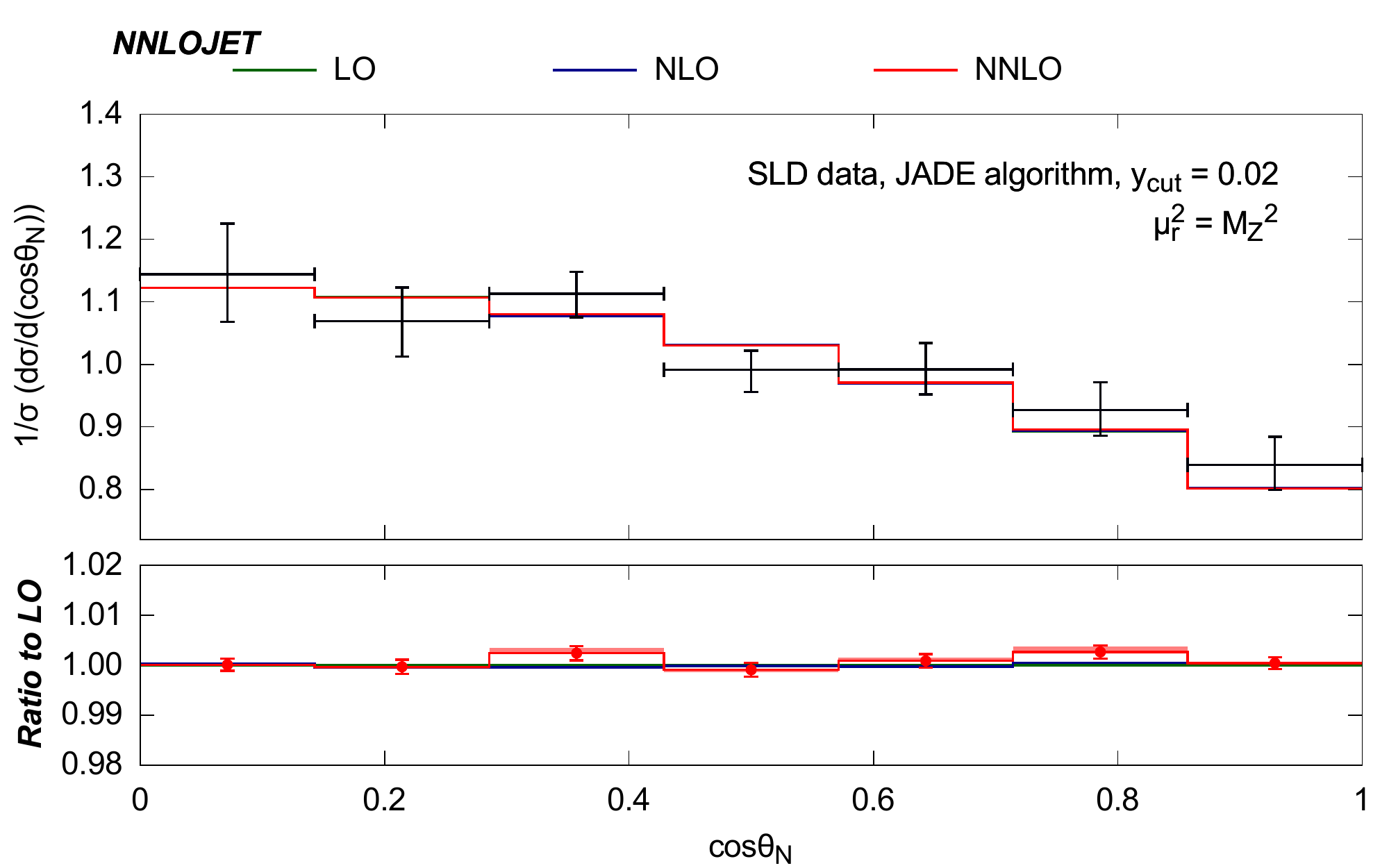}\\
  \includegraphics[width=7cm]{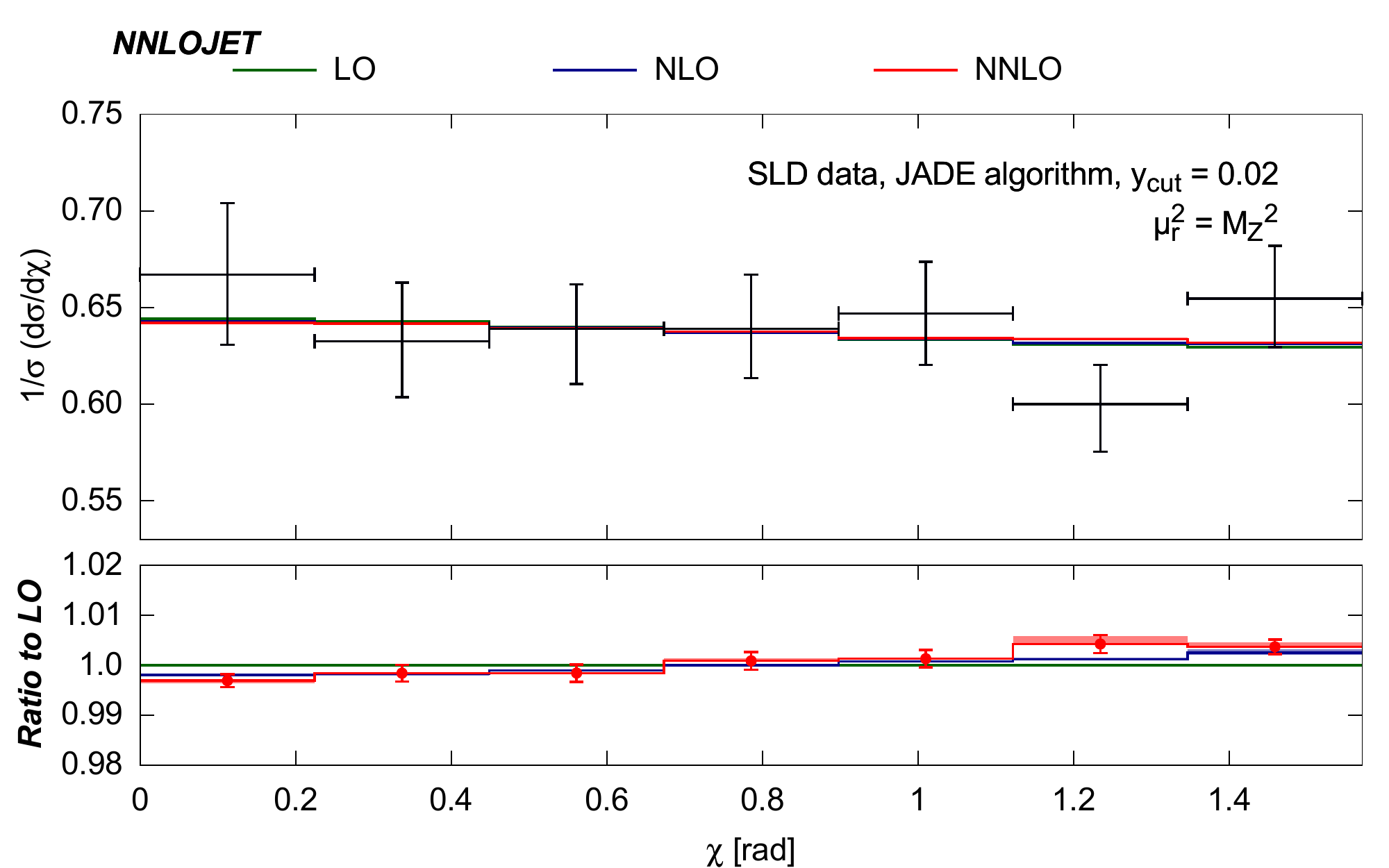}\quad
  \caption{Event orientation distributions for three-jet events (JADE algorithm, $y_{\cut}=0.02$), compared to SLD data~\protect\cite{sld}.}
   \label{fig:sld}
\end{figure}
Figure~\ref{fig:sld} displays the event orientation distributions at LO, NLO, and NNLO for exclusive three-jet events and compares them to the SLD 
data~\cite{sld}. The error bands on the NLO and NNLO predictions are obtained by varying the renormalisation scale in the strong 
coupling constant within a factor $[1/2;2]$ around the central scale $\muR=M_{\PZ}$. We also indicate the numerical integration error on the NNLO 
coefficients by a red error bar in the ratio plot. 
We observe that the perturbative corrections modify the leading-order shape of the distributions only at the level of 
four per mille at NLO and at most one per cent at NNLO. The corrections are most pronounced in $\cos(\Theta)$, where they modify the slope of the 
distribution, and are even smaller in $\chi$ and 
$\cos(\Theta_N)$.
 
 \begin{figure}[t]
\centering
\includegraphics[width=7cm]{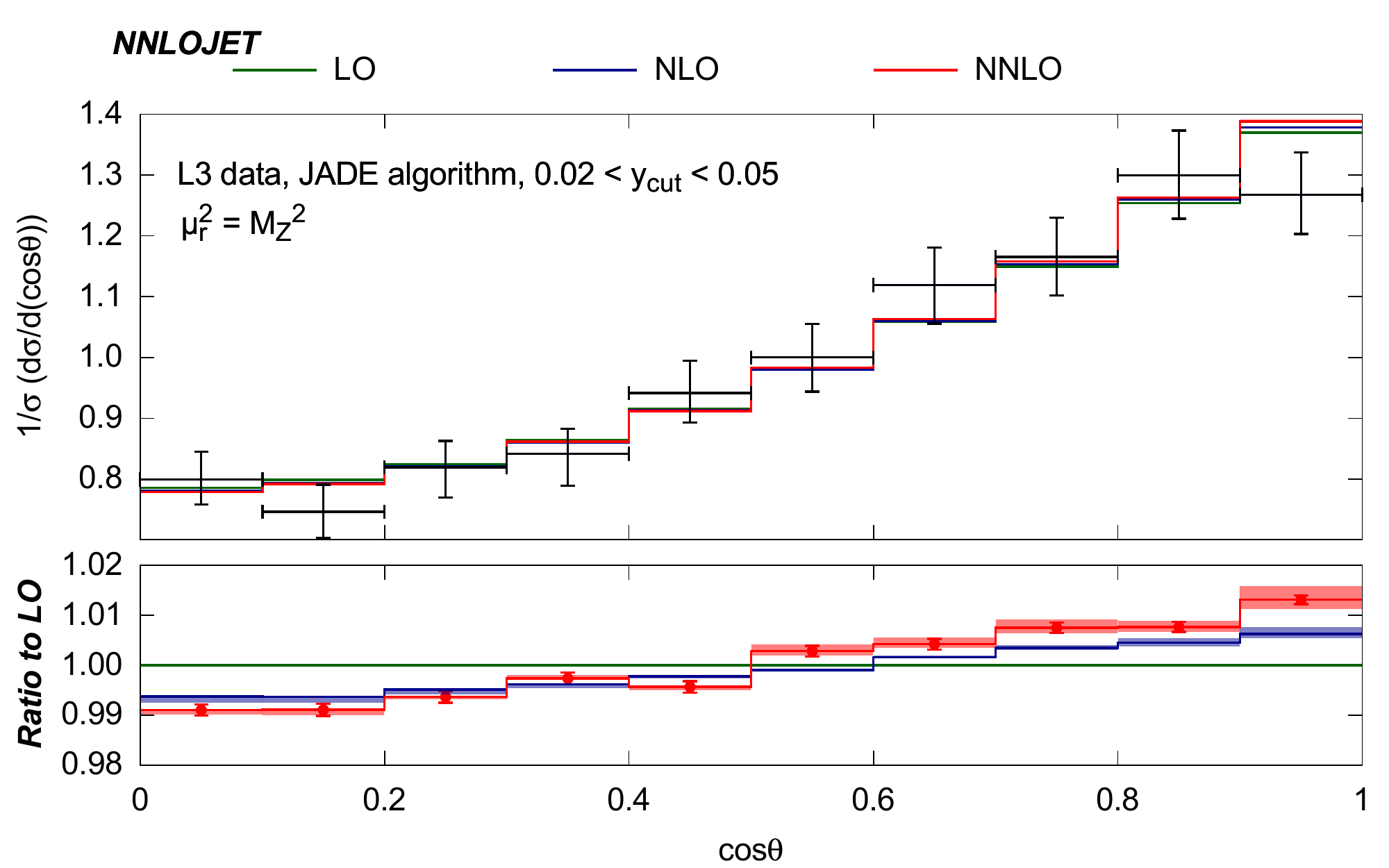}\quad
  \includegraphics[width=7cm]{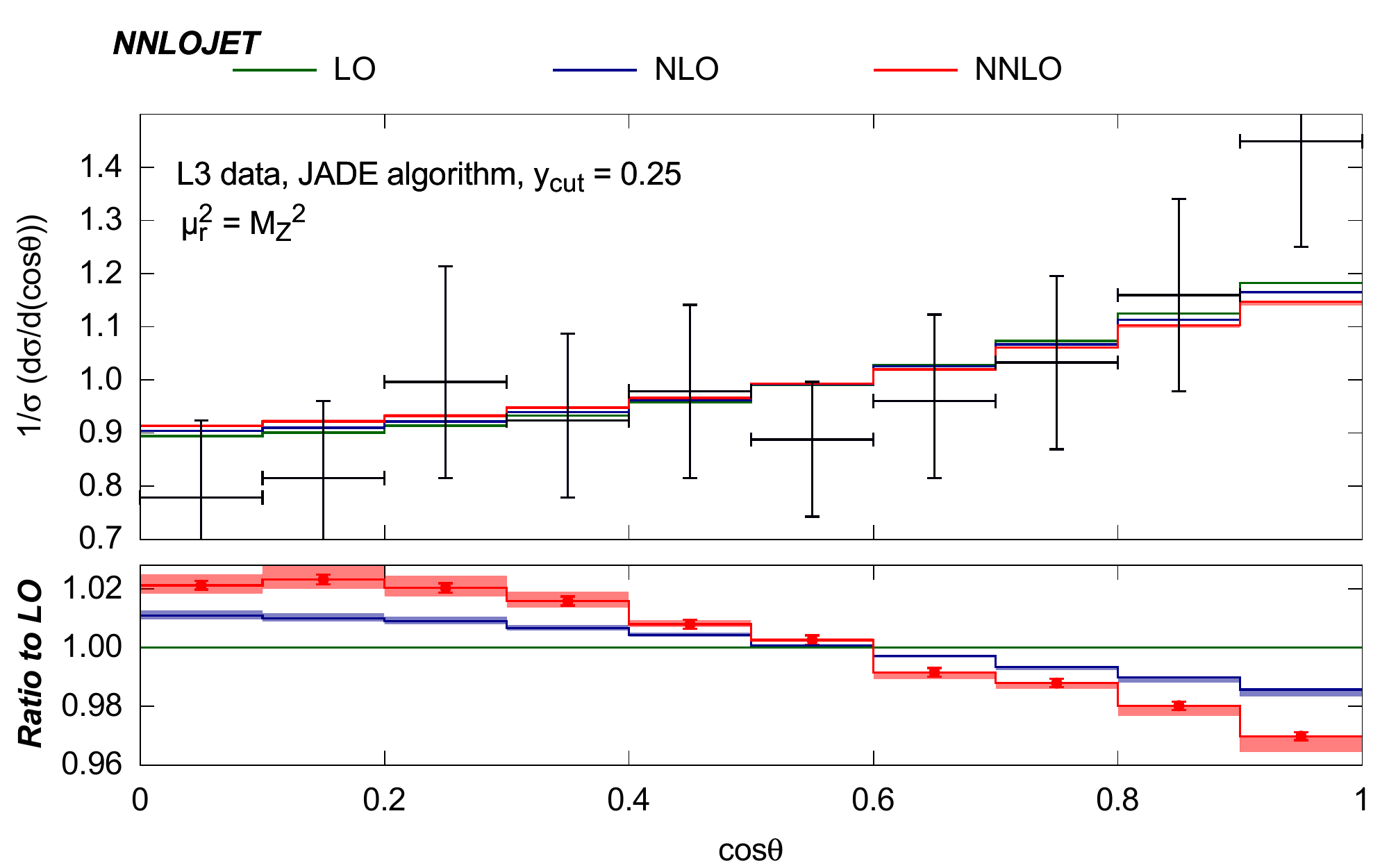}\\
  \includegraphics[width=7cm]{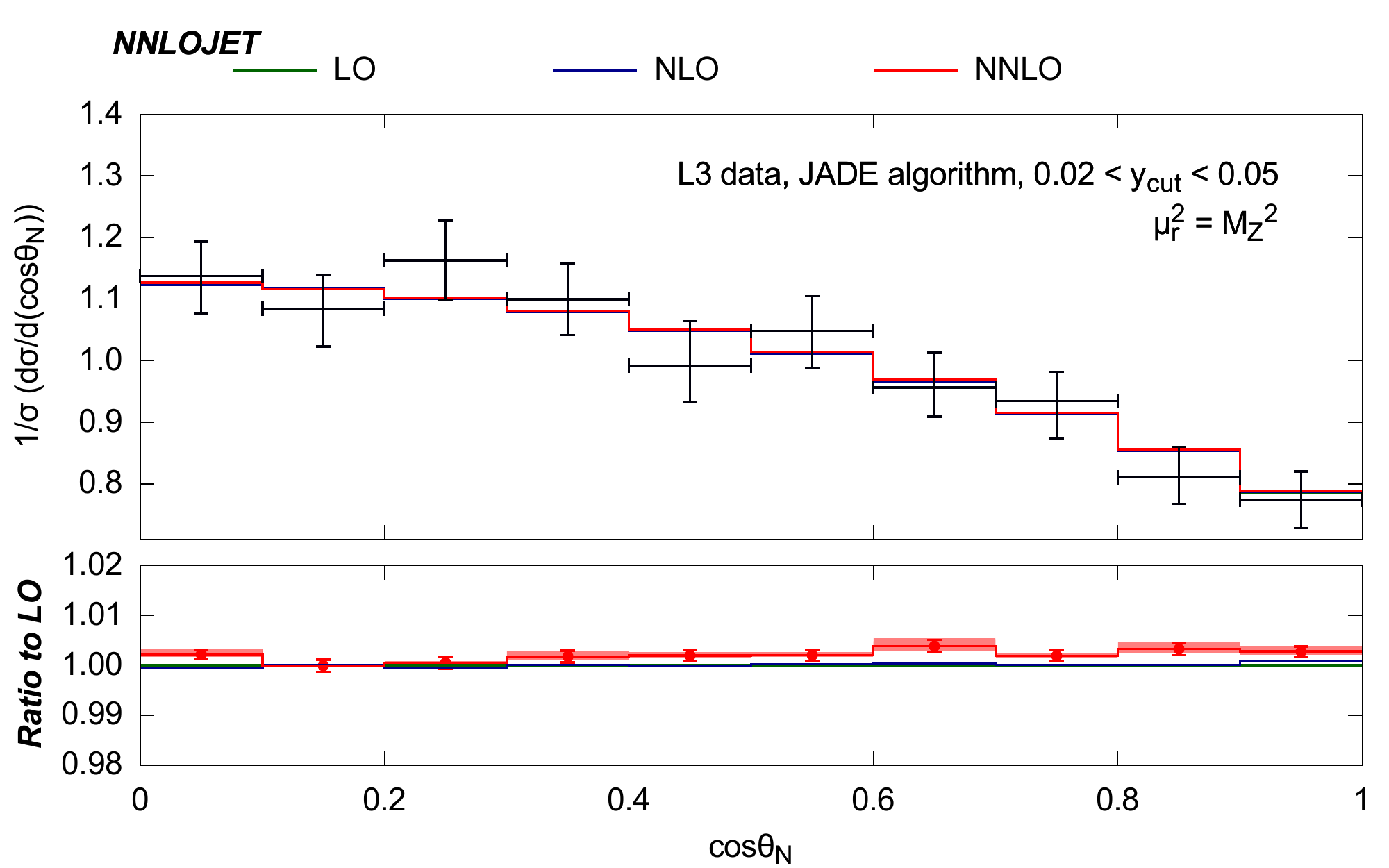}\quad
  \includegraphics[width=7cm]{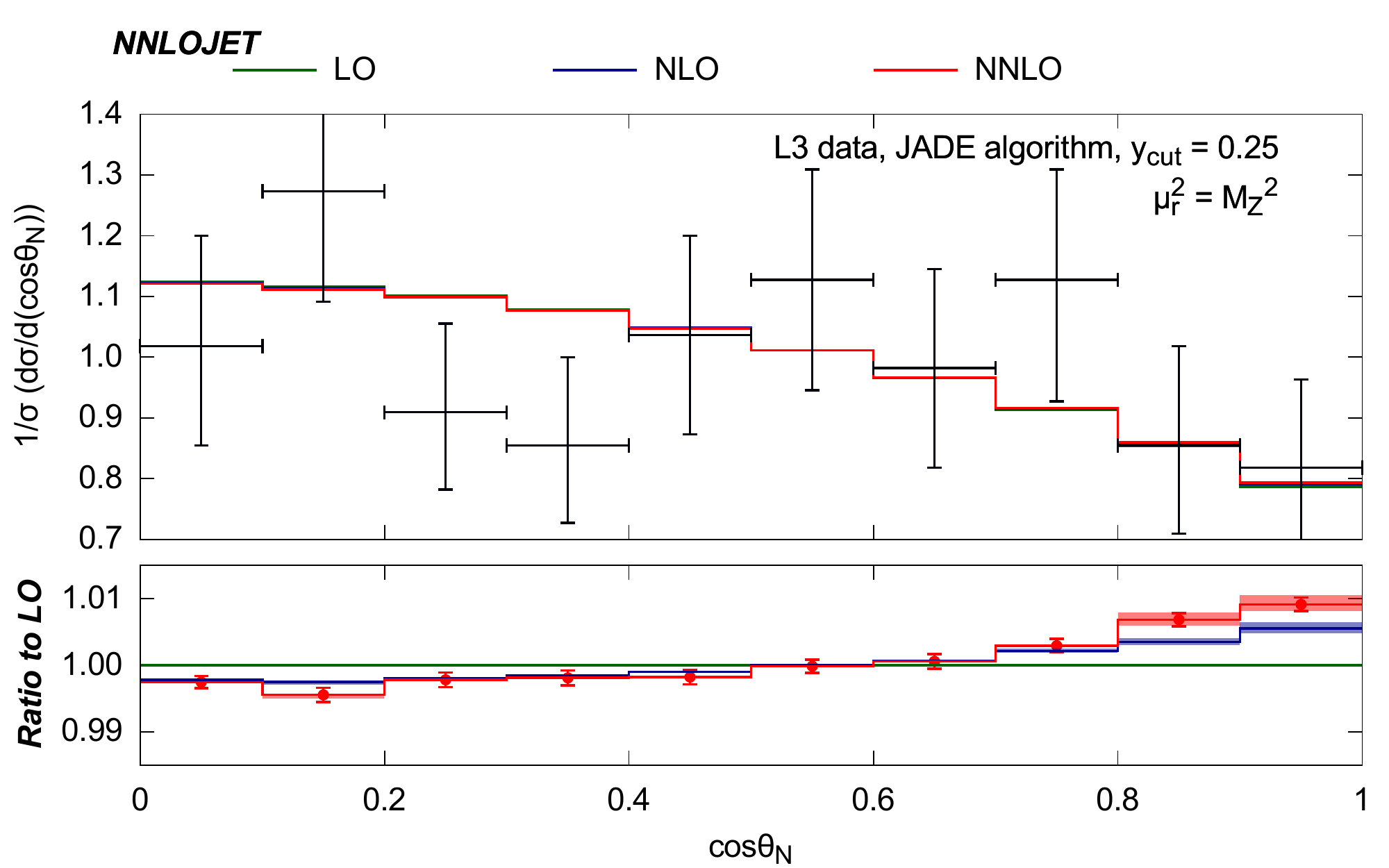}\\
\includegraphics[width=7cm]{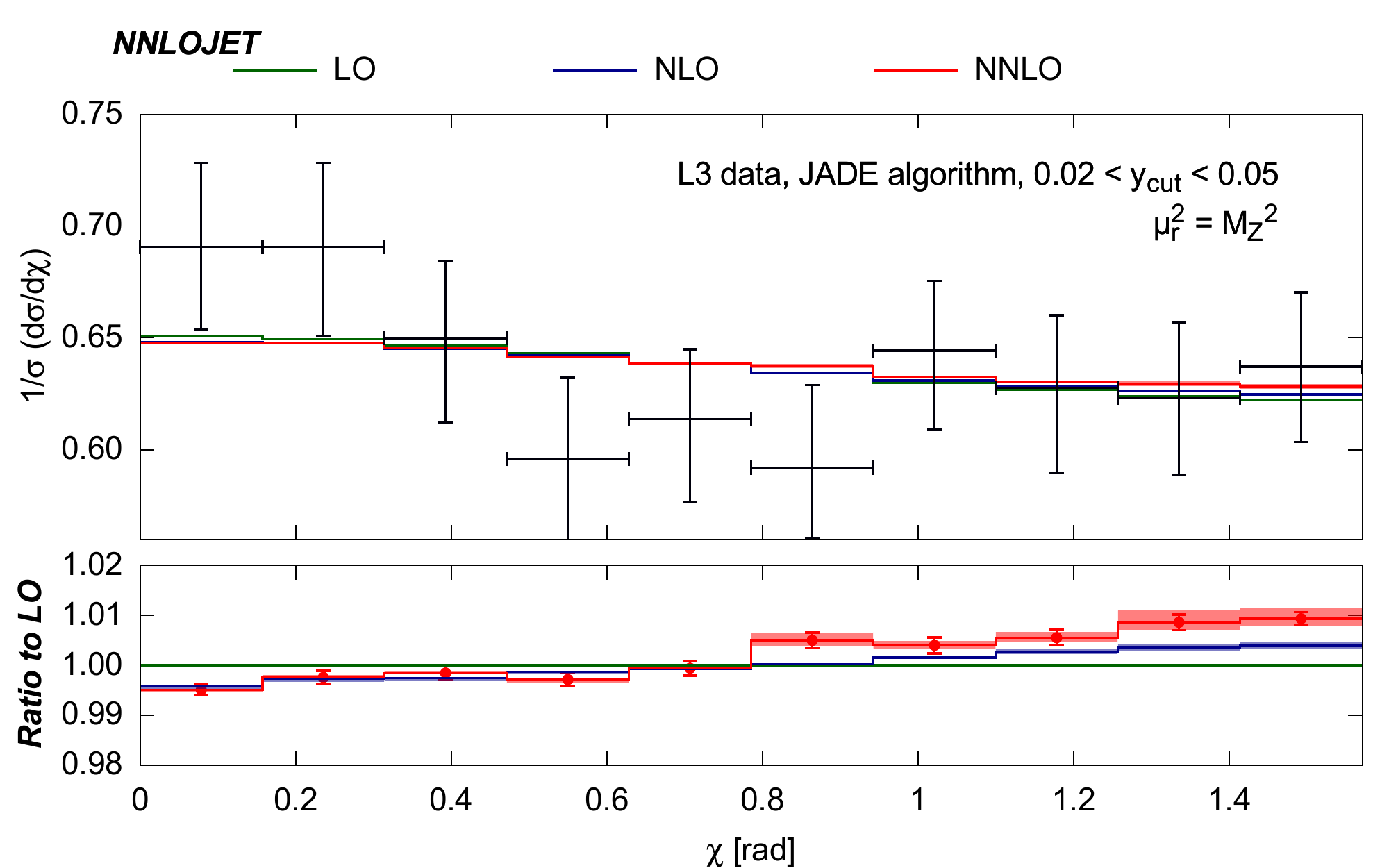}\quad
  \includegraphics[width=7cm]{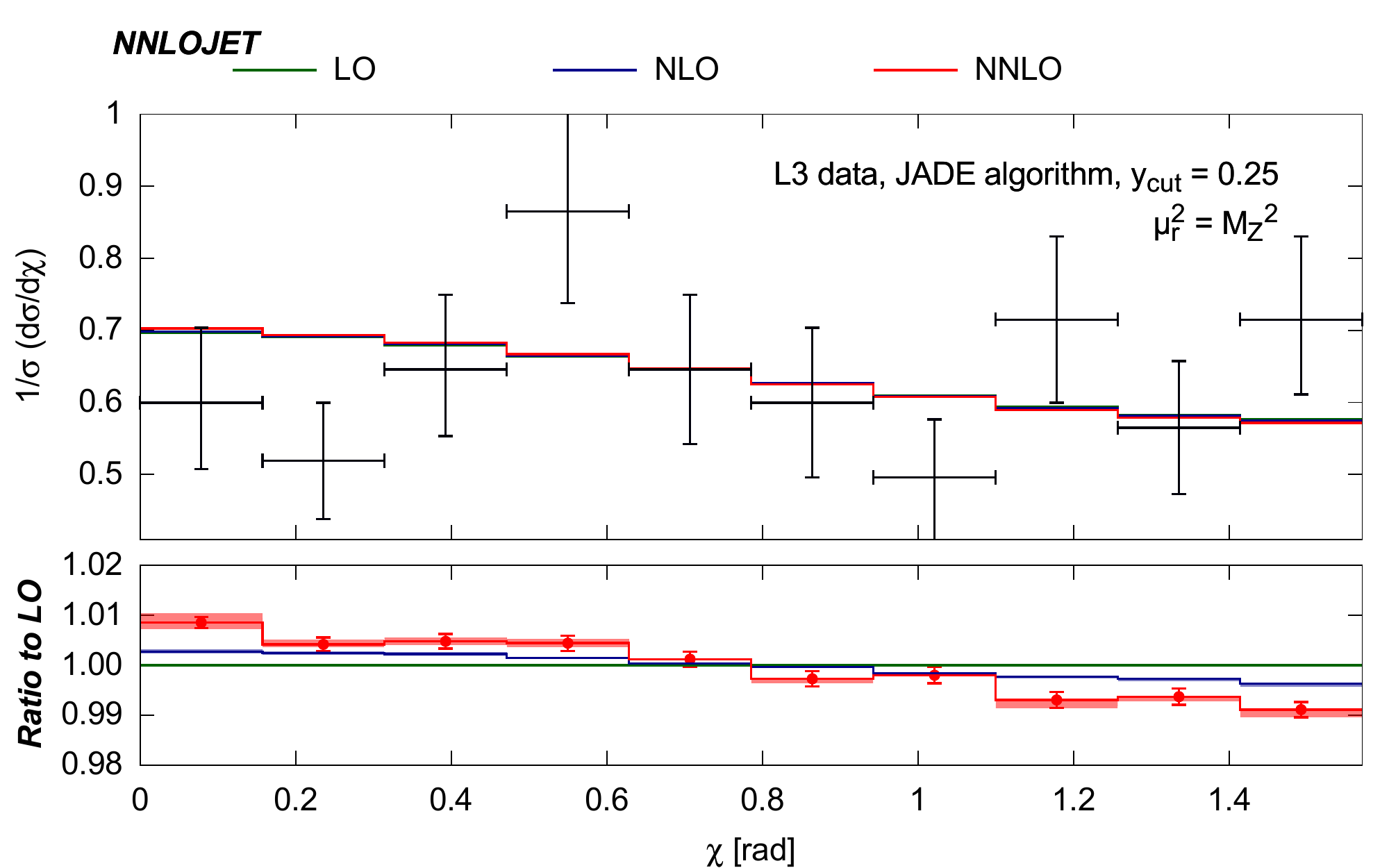}\\
  \caption{Event orientation distributions for three-jet events (JADE algorithm) compared to L3 data~\protect\cite{l3}. Left: $0.02 \leq y_{\cut}\leq 0.05$,
  right: $y_{\cut}=0.25$.}
   \label{fig:l3}
\end{figure}
The L3 experiment measured the event orientation distributions for two ranges in exclusive three-jet events (using the JADE algorithm). Results are given for two jet 
resolutions: $0.02 \leq y_{\cut}\leq 0.05$ (fine jet resolution) and $y_{\cut}=0.25$ (coarse jet resolution). 
The application of a range in $y_{\cut}$ instead of a fixed value 
is uncommon and requires further explanation: events are classified as three-jet final states if and only if they yield a three-jet final state for all values of 
$y_{\cut}$ in the interval. Since the JADE algorithm yields a monotonous increase in jet multiplicity with decreasing resolution parameter, it is sufficient 
to find a three-jet final state for both the upper and lower edge of the $y_{\cut}$ interval. 
The event orientation distributions for both values of jet resolution parameters at LO, NLO, and NNLO (with error bands and bars defined as above for SLD) 
are shown in Figure~\ref{fig:l3}, where they are compared to data from L3~\cite{l3}. For the fine jet resolution, we observe a pattern that is very similar to 
what we saw for SLD, with corrections at the level of at most one per cent throughout. For the coarse jet resolution, we observe that the corrections 
to the $\cos(\Theta)$ distribution increase to a maximum of two per cent at NNLO, and that the slope of the corrections to the $\cos(\Theta)$ and 
$\chi$ distributions is inverted compared to the fine jet resolution. 

For all distributions, we observe that the scale variation bands at NLO and NNLO do not overlap and that their size increases from NLO to NNLO. Given that the 
distributions are normalised such that they become independent of $\alphas$ at leading-order, scale variation should not be considered a good indicator 
of residual theoretical uncertainty from missing higher orders for these particular observables. The small absolute magnitude of the corrections both at NLO and NNLO 
is however a strong indicator for the perturbative stability of the event orientation distributions. 

In summary, we presented a new implementation of the NNLO QCD corrections to $\Pep\Pem \to 3\jet$ and related event-shape 
observables, using the antenna subtraction method for the cancellation of infrared singularities between real-radiation and virtual contributions. Our implementation is 
in the form of the fixed order parton-level code \NNLOJET, which can compute infrared-safe quantities using the jet definition and event selection criteria 
as used in the experimental measurements. Compared to previous implementations, we retain the full dependence on the initial-state lepton kinematics, which 
allows us to compute fiducial cross sections and event orientation distributions. The latter are particularly relevant in view of precision measurements 
of event shapes and cross sections at LEP and SLD. In these experiments, results were typically extrapolated from the actual measurements done with restricted 
detector acceptance to full $4\pi$ acceptance, using leading order multi-purpose event simulation programs. By computing the NLO and NNLO 
corrections to the event-orientation distributions, we can now quantify the impact of higher order QCD effects on these extrapolations. We find that the 
event orientation distributions are extremely robust under QCD corrections. For a fine jet resolution (where the bulk of precision QCD studies is performed), 
the corrections up to NNLO modify the distributions up to at most one per cent. By going to a more coarse jet resolution, the magnitude of the corrections increases 
slightly to two per cent, and the slopes of the corrections in some of the distributions are inverted. Our findings support the validity 
of the acceptance correction procedure applied in precision QCD studies at LEP and SLD. When aiming for per-mille level precision in QCD 
measurements at a future $\PZ$ factory, these corrections will become of relevance, and it should be considered to concentrate on measurements and 
interpretation of fiducial cross sections instead of extrapolating to full acceptance.

\section*{Acknowledgements}
  The authors thank Xuan Chen, James Currie, Aude Gehrmann--De Ridder, Juan Cruz-Martinez, Joao Pires and Tom Morgan for useful discussions and their many contributions to the \NNLOJET code. 
  The work of EWNG was performed in part at the Aspen Center for Physics, which is supported by National Science Foundation grant PHY-1066293.
   This research was supported in part by the UK Science and Technology Facilities Council,  
   by the Swiss National Science Foundation (SNF) under contracts 200020-162487 
  and CRSII2-160814, by the Research Executive Agency (REA) of the European Union under the Grant Agreement PITN-GA-2012-316704 (``HiggsTools'') and the ERC Advanced Grant MC@NNLO (340983).

\end{document}